\documentclass[review]{elsarticle}

\usepackage{lineno,hyperref}
\usepackage{amsmath,graphicx}
\usepackage{mathtools}
\usepackage{color}

\modulolinenumbers[5]

\begin{document}

\begin{frontmatter}

\title{The Glauber model and heavy ion reaction and elastic scattering cross sections}

\author[ad1]{Ajay Mehndiratta}
\author[ad2,ad3]{Prashant Shukla\corref{ca}}
  \cortext[ca]{Corresponding author}
  \ead{pshukla@barc.gov.in}
\address[ad1] {Physics Department, Indian Institute of Technology, Guwahati, India}
\address[ad2] {Nuclear Physics Division, Bhabha Atomic Research Centre, Mumbai 400085, India.}
\address[ad3] {Homi Bhabha National Institute, Anushakti Nagar, Mumbai 400094, India.}

\begin{abstract}
 We revisit the Glauber model to study the heavy ion reaction cross sections and
elastic scattering angular distributions at low and intermediate energies. 
 The Glauber model takes nucleon-nucleon cross sections and nuclear densities as
inputs and has no free parameter and thus can predict the cross sections for 
unknown systems. The Glauber model works at low energies down to Coulomb barrier 
with very simple modifications.
  We present new parametrization of measured total cross sections
as well as ratio of real to imaginary parts of the scattering amplitudes for pp and
np collisions as a function of nucleon kinetic energy.
 The nuclear (charge) densities obtained by electron scattering form factors measured 
in large momentum transfer range are used in the calculations.
  The heavy ion reaction cross sections are calculated for light and heavy systems 
and are compared with available data measured over large energy range.
The model gives excellent description of the data.
  The elastic scattering angular distributions are calculated for various systems
at different energies. The model gives good description of the data at small
momentum transfer but the calculations deviate from the data at large momentum transfer.

\end{abstract}


\end{frontmatter}

\linenumbers


\section{Introduction}
 
  The Glauber model \cite{GLAUBER} is a semi classical model picturing the nuclei
moving in a straight line trajectory along the collision direction and describes
nucleus-nucleus interaction \cite{KAROL} in terms of nucleon-nucleon (NN) interactions.
 At high energies it is used to obtain total nucleus-nucleus cross section and
geometric properties of the collisions such as the number of participants and
binary NN collisions as a function of impact parameter
[see e.g. \cite{WONG,SHUKLAarxiv2001,d'Enterria:2003qs}]. 
 At low energies, the straight line trajectory is assumed at the Coulomb distance 
of closest approach between the two nuclei \cite{CMGM,CHAGUP}. 
  The non eikonal nature of the trajectory is taken into account using
a simple prescription given in Ref.~\cite{SHUKLA1995}.
  This Coulomb Modified Glauber Model (CMGM) has been widely used in
the literature \cite{WARNER,AHMAD,CAI,PNU}. The work in Ref.~\cite{SHUKLAprc2003}
presents a systematic calculation of reaction cross section using CMGM. 
  The Glauber model has no free parameter and thus has a variety of applications. 
It is used to extract the radii of unstable nuclei from measured total or
reaction cross sections \cite{Alkhazovi:2011ty,TANIHATA, Horiuchi:2006ga}.
  The Glauber model formalism together with the measured cross section 
is frequently used as a tool to test various forms of relativistic 
mean field densities \cite{PATRA2009, Sharma:2013nba, Chauhan:2014uoa}.
 It is also a useful tool to study the shape deformation of nuclei \cite{Hassan:2009dc,Christley:1999jat}.
 The Glauber approach is similar to microscopic optical model approach which is
used to study various nuclear reaction mechanisms \cite{Lukyanov:2015aya}.
  The elastic scattering data are mostly interpreted in terms of optical 
model potential where the real part is commonly taken as double folding potential.
  Such formalism uses an imaginary potential with three free parameters and 
reproduces the diffractive patterns up to large angles as shown in the work of 
Ref.~\cite{DTKhao:2000npa} for the elastic scattering of   
$^{16}$O + $^{16}$O at incident energies ranging from 124 to 1120 MeV.
  There are numerous attempts to explain the elastic scattering angular distributions
of light nuclei using the Glauber model \cite{ElGogary:2000cv}. 
 To have a better agreement with the data, the NN scattering amplitude is modified  
to include phase variation \cite{ElGogary:1998gu}. 
 The isospin effects in NN scattering process have small impact on the 
cross sections as shown in the work of Ref.~\cite{Sammarruca:2011nc} which
calculates the reaction cross section at 
intermediate energy to study the medium effects on NN scattering cross sections.
  Ref.~\cite{Ahmad:2004alvi} describes the elastic scattering angular distributions of 
$^{16}$O + $^{16}$O  and $^{12}$C + $^{12}$C using an NN phase shift function having 
three free parameters.
   A detailed study in the Ref.~\cite{Gibbs:2012yd} presents a Monte Carlo Glauber model 
calculations of angular distributions of elastic scattering of $\alpha$ on   
light and heavy nuclei. The Monte Carlo approach includes the geometric fluctuations
but is expected to give similar results as optical Glauber model at low energies.
On comparison with the data this work concludes that the 
angular distributions can be predicted only up to certain angles. To get a better 
agreement at higher angles one may require more parameters as in Ref.~\cite{Ahmad:2004alvi}.

In view of the importance and wide applicability of the model, we extend reaction cross 
section study of work in Ref.~\cite{SHUKLAprc2003} for many more systems and collisions 
energies. We also calculate elastic scattering angular distributions, a study similar to 
the work of Ref.~\cite{Gibbs:2012yd} but for many more systems.
The nucleon-nucleon  cross sections  $\sigma_{nn}$, $\sigma_{pp}$ and
$\sigma_{np}$ are the most important inputs in the calculations.
  We present simple parametrizations for the total cross sections
as well for the ratio of real to imaginary parts of the scattering amplitudes for pp and
np collisions using a large set of measurements and make a comparison with
those available in the literature \cite{CHAGUP,Bertulani:2010kk}.
  The nuclear (charge) densities obtained by electron scattering form factors measured 
  in large momentum transfer range are used in the calculations \cite{JAGER1974,JAGER1987}.
  For few systems we use 
three parameter Fermi density (3pF) in contrast to two parameter Fermi (2pF) density 
and Gaussian densities used in previous studies. 
  The center of mass correction which is important for light systems has also been taken 
into account \cite{Franco1987}.
  The reaction cross section and the elastic scattering angular distributions are obtained 
at many energies and are compared with the data to test the reliability of the model 
and the input parameters for many cases of stable nuclei.

\section{The Glauber Model}
   The Glauber model gives the probability for occurrence of a nucleon-nucleon collision 
when the nuclei $A$ and $B$ collide at an impact parameter  ${\bf b}$ relative to 
each other which is determined to be \cite{WONG,SHUKLAarxiv2001}
\begin{equation}\label{wong}
T(b) \bar \sigma_{NN} = \int \rho_A^z({\bf b}_A) d{\bf b}_A \,
 \rho_B^z({\bf b}_B) d{\bf b}_B \,
 t({\bf b-b_A+b_B}) \,\, \bar \sigma_{NN}.
\end{equation}
\noindent 
 Here, $\rho_A^z({\bf b}_A)$ and $\rho_B^z({\bf b}_B)$
are the z-integrated densities of projectile and target nuclei
respectively. $t({\bf b}) d {\bf b}$ is the probability for having
a nucleon-nucleon collision within the transverse area element
$d {\bf b}$ when one nucleon approaches at an impact parameter
${\bf b}$ relative to another nucleon.
All these distribution functions are normalized to one.
Here $\bar \sigma_{NN}$ is the average total nucleon nucleon cross section.

  The total reaction cross section $\sigma_R$ can be written as
\begin{align*}
  \sigma_{R} &= 2 \pi \int b db \left( 1 - |S(b)|^2 \right), \\
  &= \frac{\pi}{k^{2}} \, \sum_{l=0}^\infty \, (2l+1)(1-|S_{l}|^{2}).
\end{align*}
  The scattering matrix $S_{l}$ or $S(b)$ where $bk=(l+1/2)$ is given by 
\begin{eqnarray}\label{sb2}
S(b)= \exp\left(i\chi (b)\right).
\end{eqnarray}
 The Glauber phase shift $\chi(b)$  can be written as 
\begin{eqnarray}\label{rf8}
\chi (b) ={1\over 2}\bar \sigma_{NN} (\bar \alpha_{NN}+i) \, AB \,\,T(b).
\end{eqnarray}
  Here, $\bar \alpha_{NN}$  is the ratio of real to imaginary part of NN scattering 
amplitude which does not appear in the calculations of reaction cross section but 
is important for elastic scattering angular distribution. 

 In momentum space, $T(b)$ is derived as \cite{SHUKLAprc2003}
\begin{eqnarray}\label{tbmom}
T(b) = {1\over 2\pi}
       \int J_0(qb) S_{A}({\bf q}) S_{B}(-{\bf q})  f_{NN}(q) q dq.
\end{eqnarray}
Here, $S_{A}(q)$ and  $S_{B}(-q)$ are the Fourier transforms of the nuclear densities 
and $J_0(qb)=1/2\pi\int \exp(-qb \cos \phi) d\phi$ is the
cylindrical Bessel function of zeroth order.
The function $f_{NN}(q)$ is the Fourier transform of the
profile function $t({\bf b})$ and gives the $q$ dependence of
NN scattering amplitude.
 The profile function $t({\bf b})$ for the NN scattering can be
taken as delta function if the nucleons are point particles.
In general, it is taken as a Gaussian function of width $r_0$ as
\begin{equation}
t({\bf b}) = {\exp(-b^2/2r_0^2) \over (2\pi r_0^2)}.
\end{equation}
Thus,
\begin{eqnarray}
f_{NN}(q) = \exp(-r_0^2 q^2/2).
\end{eqnarray}
  Here, $r_0$ is the range parameter and has a weak dependence on 
energy (see for discussions \cite{AHMAD}). For the present work,
we use $r_0=0.6$ fm, which is guided by the previous
studies in the same energy region from Refs. \cite{SHUKLAprc2003,CHAGUP}.

  In the presence of Coulomb field, the non eikonal trajectory around the
Coulmob distance of closest approach $r_c$ is represented by
$r^2 = r_c^2 + (C+1) z^2$ \cite{SHUKLA1995} where $r_{c}$ and the factor $C$
are given by 
\begin{eqnarray}
  r_{c} = (\eta + \sqrt{\eta^{2} + b^{2}k^{2}})/k,
\end{eqnarray}
\begin{eqnarray}
  C = \frac{\eta}{kb^{2}}r_{c}.
\end{eqnarray}
Here, $\eta = Z_{P}Z_{T}e^{2}/\hbar v$  is the dimensionless Sommerfield parameter.
In the Coulomb Modified Glauber Model (CMGM) the Eq.~\ref{rf8} is modified as

\begin{eqnarray}\label{rf8CMGM}
\chi (b) ={1\over 2}\bar \sigma_{NN} (\bar \alpha_{NN}+i) \, AB \,\,T(r_c)/\sqrt{C+1}.
\end{eqnarray}


\section{Elastic Scattering Cross Section}
 The nucleus-nucleus differential elastic cross section as a function of center of mass 
angle $\theta$ is given by 
\begin{eqnarray}\label{dfecs}
  {\frac{d\sigma_{el}}{d\Omega}} = |f(\theta)|^{2},
\end{eqnarray}
where $f(\theta)$ is the sum of Coulomb and nuclear scattering amplitudes.
\begin{eqnarray}\label{F=Fc+Fn}
  f(\theta) = f_{C}(\theta) + f_{N}(\theta).
\end{eqnarray}
For identical systems, the LHS of Eq.~\ref{dfecs} is replaced
by $|f(\theta) + f(\pi-\theta)|^{2}$.
The Coulomb scattering amplitude is given by 
\begin{eqnarray}\label{Fc}
  f_{C}(\theta) = A_C \, e^{i\phi_{C}},
\end{eqnarray}
where $A_C = -{\frac{\eta}{2k}}cosec^{2}{\frac{\theta}{2}}$, \, $\phi_{C} = 2\sigma_{0} - 2\eta \ln \bigg(\sin{\frac{\theta}{2}}\bigg)$
and the nuclear scattering amplitude is 

\begin{eqnarray}\label{Fn}
  f_{N}(\theta) = {\frac{1}{2ik}}\sum_{l=0}^\infty (2l+1)(e^{2i\sigma_{l}})(S_{l}-1)P_{l}(\cos\theta).
\end{eqnarray}
Here, $\sigma_{l+1}(\eta) = \sigma_{l}(\eta) + tan^{-1}(\frac{\eta}{l+1})$ and
$\sigma_{0}$ can be assumed to be 0. The nuclear scattering amplitude can be written as 
$S_{l} = \exp[i(\chi_{R} + i\chi_{I})]$  and thus 
$f_{N}(\theta)$ is simplified to 

\begin{eqnarray}\label{NrNi}
  f_{N}(\theta) =  \frac{1}{2k} \sum_{l=0}^\infty \, (2l+1) \, P_{l}(\cos\theta)\, (e^{2i\sigma_{l}}) \, [N_{R} + iN_{I}] ,
\end{eqnarray}
where $N_{R} = [e^{-\chi_{I}} \, \sin\chi_{R}]$ and $N_{I} = [1 - e^{-\chi_{I}} \, \cos\chi_{R}]$.

\section{The Nuclear Densities}
\begin{table}
  \caption[]{Density parameters of nuclei used in the present work. The parameters $\alpha$
    and $a$ correspond to MHO density, the parameters $d$ and $c$ correspond to 2pF/3pF
    densities and $\omega$ corresponds to 3pF density. $q$-range is measured momentum
  transfer range.} 
\label{DensityParameters}
\begin{tabular}{l l l l l l l l} 
\hline
\hline
  Element     &   Form  & $d$/$\alpha$ & $R_{rms}$ &  $c$/$a$ &  $w$   &  $q$-range   &  Ref.      \\ 
              &         &    (fm)      &   (fm)   &   (fm)   &        &              &            \\
\hline
$^{12}_{6}$C    &    MHO     &   1.247(18)      &  2.460      &   1.649(8)   &       & 1.05-4.01  & \cite{JAGER1974}    \\ 
$^{16}_{8}$O    &    HO      &   1.517      &  2.674   &  1.805(15)   &       & 0.58-0.99  &  \cite{JAGER1974}   \\      
$^{28}_{14}$Si  &    2pF     &   0.542(16)      &  3.138   &  3.106(30)   &        & 0.41-2.02  & \cite{JAGER1974}    \\     
$^{40}_{20}$Ca  &    3pF     &   0.584      &  3.486   &  3.669   &  -0.102 & 0.49-3.37  & \cite{JAGER1974}    \\ 
$^{90}_{40}$Zr  &    2pF     &   0.55       &  4.274   &  4.712   &        &     -        & \cite{SHUKLAprc2003}  \\ 
$^{208}_{82}$Pb &    2pF     &   0.549(8)       &  5.521  &  6.624(35)  &       & 0.22-0.88  & \cite{JAGER1974}    \\ 
\hline 
\hline
\end{tabular} \\
\end{table}

  The nuclear densities of the two nuclei are the most important inputs in the model.
 We can calculate the Fourier transform for 
any given density form $\rho(r)$ to be used in Eq.~(\ref{tbmom}) as follows
\begin{eqnarray}
  S(q) = 4 \pi \int j_{0}(qr) \, \rho(r) \, r^{2} dr
\end{eqnarray}
  Here, $j_0(qr)$ is the spherical Bessel function of order zero.
 The nuclear densities are obtained by fitting electron nucleus scattering form factors measured 
in a momentum transfer range \cite{JAGER1974,JAGER1987}.
  For light nuclei such as $^{6}$Li, $^{12}$C and $^{16}$O,
we use the Modified Harmonic Oscillator (MHO) density with correction for center of mass motion \cite{SHUKLAprc2003}.
For heavier nuclei such as $^{28}$Si, $^{90}$Zr and $^{208}$Pb we use two parameter Fermi (2pF) density.
 We also use the three parameter Fermi (3pF) density for nuclei such as $^{40}$Ca for which 2pF density is not 
given for wide q-range of measured form factor.
 The mean radius, $c$ for $^{90}$Zr has been calculated using the formula given in Ref.~\cite{SHUKLAprc2003}. 

The MHO density form is given by 

\begin{eqnarray} \label{rhoHO}
  \rho(r) = \rho_{0}\bigg(1+\alpha \frac{r^{2}}{a^{2}}\bigg)\,\exp\bigg(-\frac{r^{2}}{a^{2}}\bigg) \,, \,\,\,\, 
\rho_{0} = \frac{1+1.5\alpha}{(\sqrt{\pi}\,a)^{3}}.
\end{eqnarray}
The 2pF density is given by

\begin{eqnarray} \label{rho2pF}
  \rho(r) = \frac{\rho_{0}}{1+\exp(\frac{r-c}{d})} \,, \,\,\,\,\, 
\rho_{0} = \frac{3}{4\pi c^{3}[1 + \frac{\pi^{2}d^{2}}{c^{2}}]}
\end{eqnarray}
 and the 3pF density is given by 

\begin{eqnarray} \label{rho3pF}
  \rho(r) = \frac{\rho_{0}(1+\frac{wr^{2}}{c^{2}})}{1+\exp(\frac{r-c}{d})}.
\end{eqnarray}
The parameters for different nuclei used in the present work are given 
in Table~\ref{DensityParameters}.


\section{NN scattering parameters}
  
  The average NN scattering parameter $\bar \sigma_{NN}$ is obtained in terms of pp cross section 
$\sigma_{pp}$ and np cross section $\sigma_{np}$ averaged over proton numbers ($Z_{P}$, $Z_{T}$) 
and neutron numbers ($N_{P}$, $N_{T}$) of projectile and target respectively as 

\begin{eqnarray}
  \bar \sigma_{NN} = \frac{N_{P}N_{T}\sigma_{nn} + Z_{P}Z_{T}\sigma_{pp} + (Z_{P}N_{T} + N_{P}Z_{T})\sigma_{np}}{A_{P}A_{T}} .
\end{eqnarray}
  The parametrized forms of $\sigma_{pp}$ and $\sigma_{np}$ are available in literature \cite{CHAGUP,Bertulani:2010kk}.
The cross section $\sigma_{pp}$ is assumed to be the same as $\sigma_{pp}$.
We obtain new parametrization using the data from Particle Data Group \cite{PDG} which are given 
in terms of proton lab kinetic energy $E$ as follows

\begin{eqnarray} \label{sigpp}
  \sigma_{pp} = \left\{
  \begin{array}{ll}
     -5.32 + {3017.0 \over E}                           &   : 9\leq E ({\rm MeV}) \leq90 \\
     31.74 -0.0628\,E + 1.22\times 10^{-4}\,E^2  &  : 90\leq E ({\rm\, MeV}) \leq700 
  \end{array}
  \right. ,
\end{eqnarray}
      
\begin{eqnarray}\label{signp}
  \sigma_{np} = \left\{
  \begin{array}{ll}
     -1128.0 - {863.0\over E} + 61.5\sqrt{E} + {6170.0 \over \sqrt{E}}  &  : 0.45\leq E ({\rm MeV}) \leq70 \\
     -8.54 + {9956.0 \over E} + 1.64\sqrt{E} - {321.0 \over\sqrt{E}}     &          : 70\leq E ({\rm MeV}) \leq700 
  \end{array}
\right. 
\end{eqnarray}

\begin{figure}
  \includegraphics[width=0.49\textwidth]{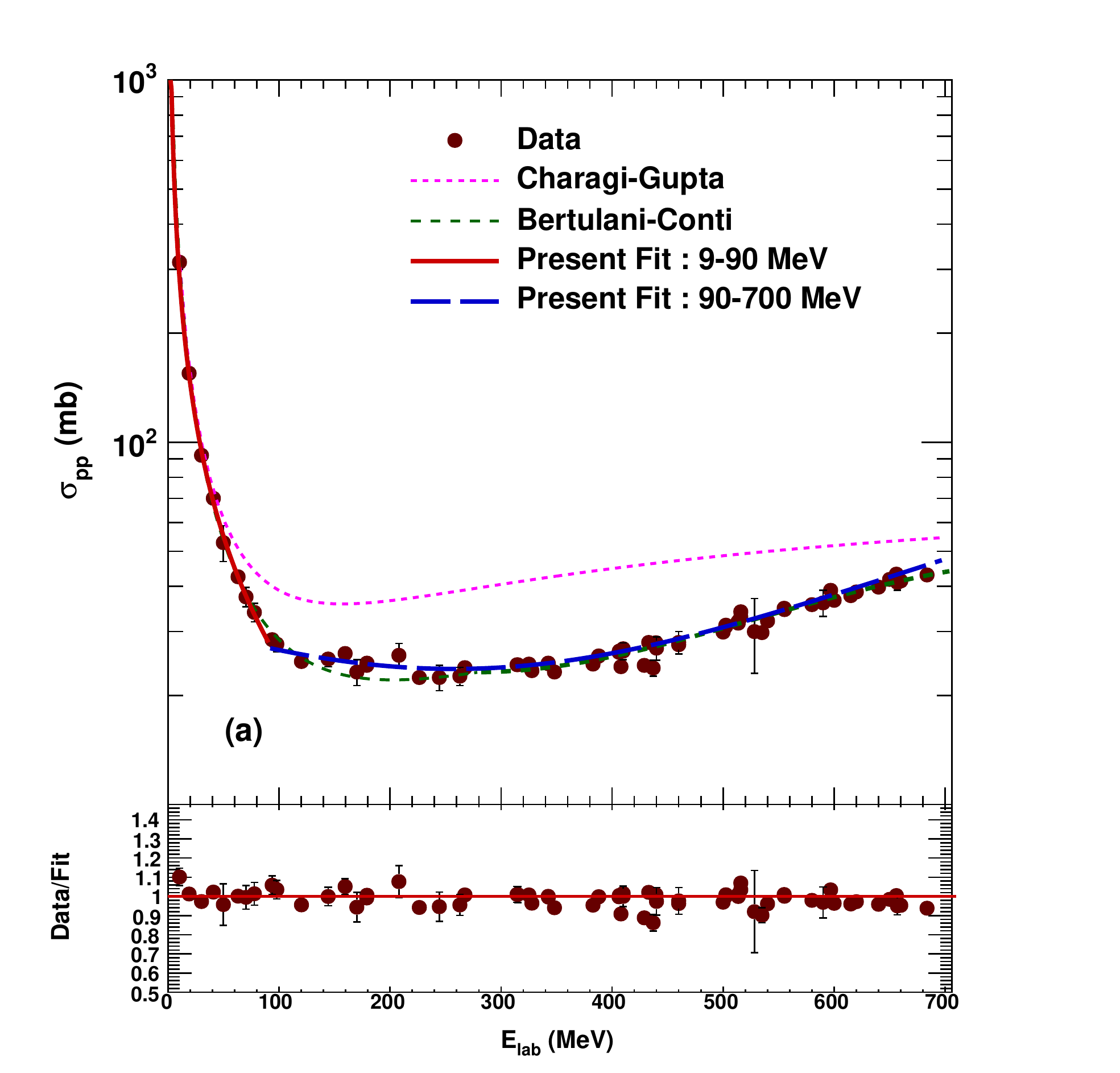}
  \includegraphics[width=0.49\textwidth]{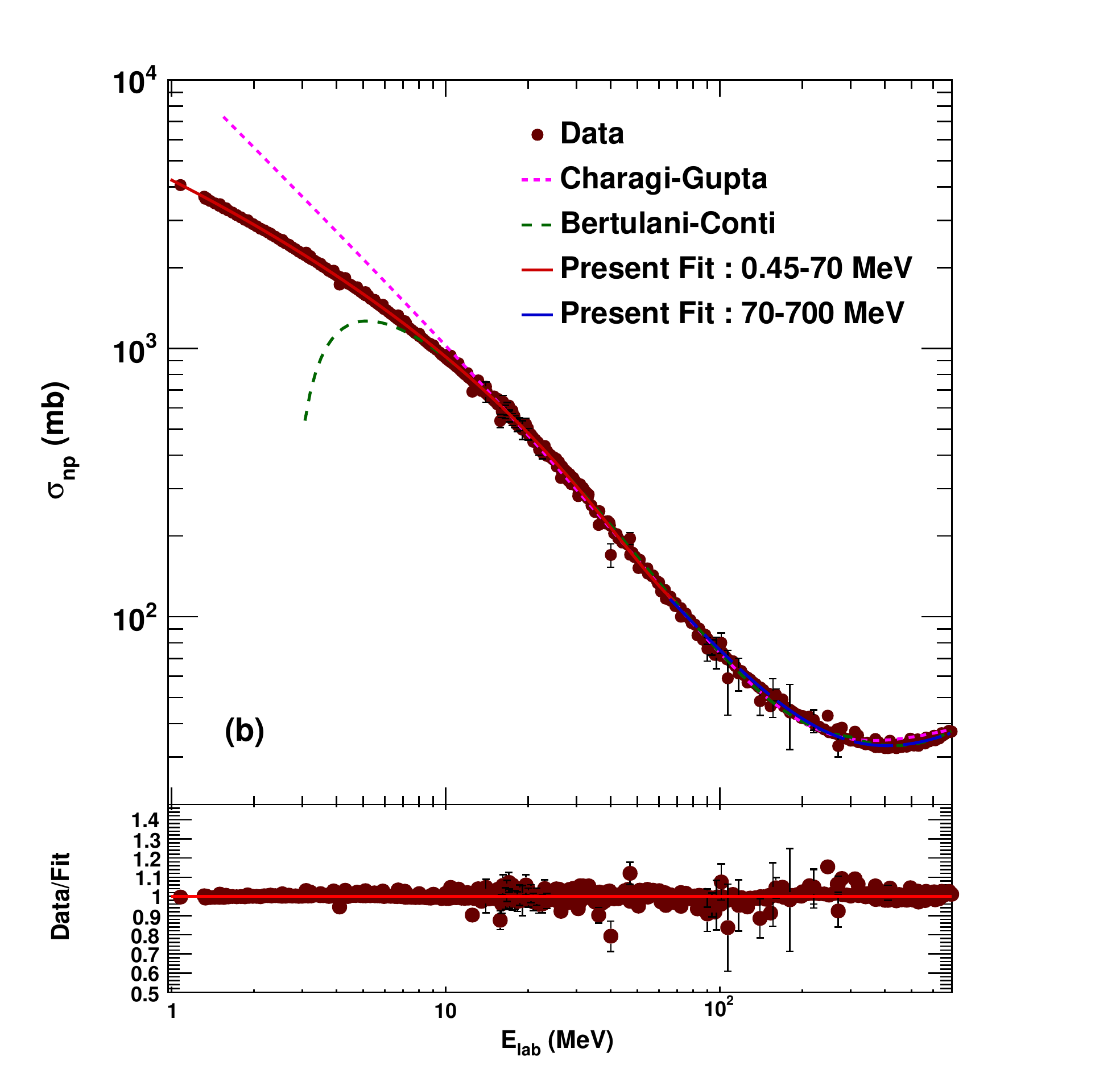}
  \caption{(a) proton-proton cross section and (b) neutron-proton cross section data 
as a function of lab kinetic energy \cite{PDG} fitted with the functions given by Eqs.~(\ref{sigpp}) 
and (\ref{signp}) along with the fits which are given in references~\cite{CHAGUP,Bertulani:2010kk}.}
  \label{pp_np_cross_section}
\end{figure}

 The errors on the parameters in Eqs.~(\ref{sigpp}) and (\ref{signp}) are 1-3 \%.
Figure~\ref{pp_np_cross_section} shows the NN cross section data \cite{PDG} 
as a function of lab kinetic energy fitted with the functions given by Eqs.~(\ref{sigpp}) 
and (\ref{signp}) along with the fits given in references~\cite{CHAGUP,Bertulani:2010kk}.
The data/fit graphs are shown for the present parametrizations.
The Charagi-Gupta parametrization for $\sigma_{pp}$ is good upto 50 MeV and that
for $\sigma_{np}$ is good above 10 MeV. The Bertulani-Conti parametrization of $\sigma_{pp}$
differs with our parametrizations in the proton energy range 120-300 MeV and
their parametrization of $\sigma_{np}$ cannot be extrapolated below 8 MeV.

  The average $\bar \alpha_{NN}$ is also calculated using $\alpha_{pp}$, $\alpha_{nn}$ and $\alpha_{np}$ as follows
\begin{eqnarray} \label{alphaNN}
  \bar \alpha_{NN} = \frac{N_{P}N_{T}\alpha_{nn}\sigma_{nn} + Z_{P}Z_{T}\alpha_{pp}\sigma_{pp} + 
(Z_{P}N_{T} + N_{P}Z_{T})\alpha_{np}\sigma_{np}}{\sigma_{NN}A_{P}A_{T}} .
\end{eqnarray}
 The parametrizations for quantities $\alpha_{pp}$ and $\alpha_{np}$ 
as a function of lab kinetic energy are obtained 
using the data of phase shift analysis given in Ref.~\cite{WGREIN} which is 
in accordance with the experimental data. It is assumed that $\alpha_{nn}$ = $\alpha_{pp}$. 
The present parametrizations are obtained for proton lab energy $E$ between 7 to 260 MeV
given as

\begin{eqnarray}\label{alpp}
  \alpha_{pp} = a_{p} + b_{p}E + c_{p}E^{2} + d_{p}E^{3}  \,\,\,\, E \,\, ({\rm in \, MeV}),
\end{eqnarray}
where $a_{p} = 0.435$, $b_{p} = 3.202 \times 10^{-2}$, $c_{p} = -2.287 \times 10^{-4}$ and $d_{p} = 4.134 \times 10^{-7}$ and
\begin{eqnarray}\label{alnp}
  \alpha_{np} = a_{n} + b_{n}E + c_{n}E^{2} + d_{n}E^{3}  \,\,\,\, E \, ({\rm in \, MeV}), 
\end{eqnarray}
where $a_{n} = -0.3695$, $b_{n} = 3.211 \times 10^{-2}$, $c_{n} = -2.117 \times 10^{-4}$ and $d_{n} = 3.672 \times 10^{-7}$. The errors on these parameters are 7-10 \%. 

\begin{figure}
  \includegraphics[width=0.49\textwidth]{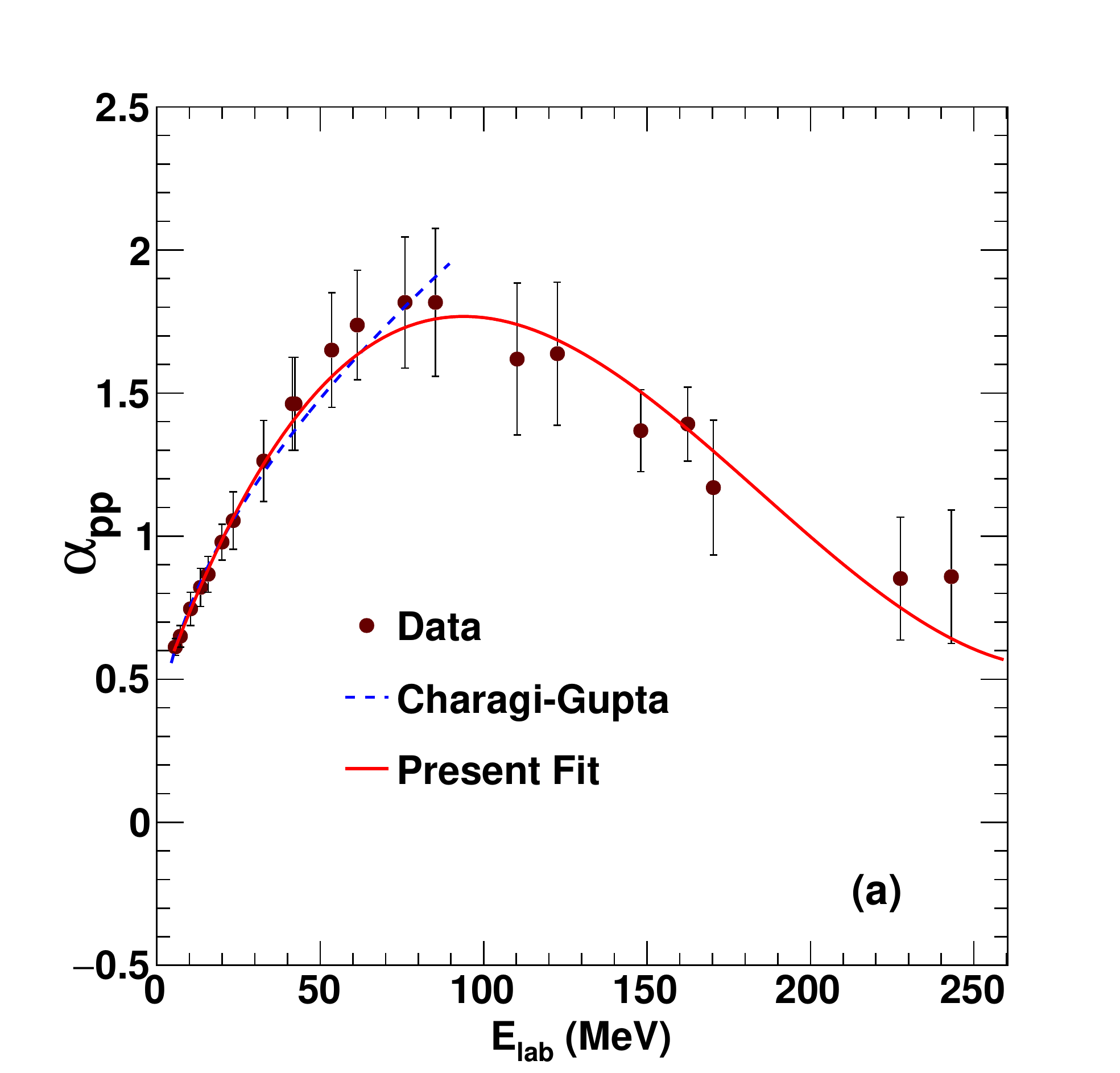}
  \includegraphics[width=0.49\textwidth]{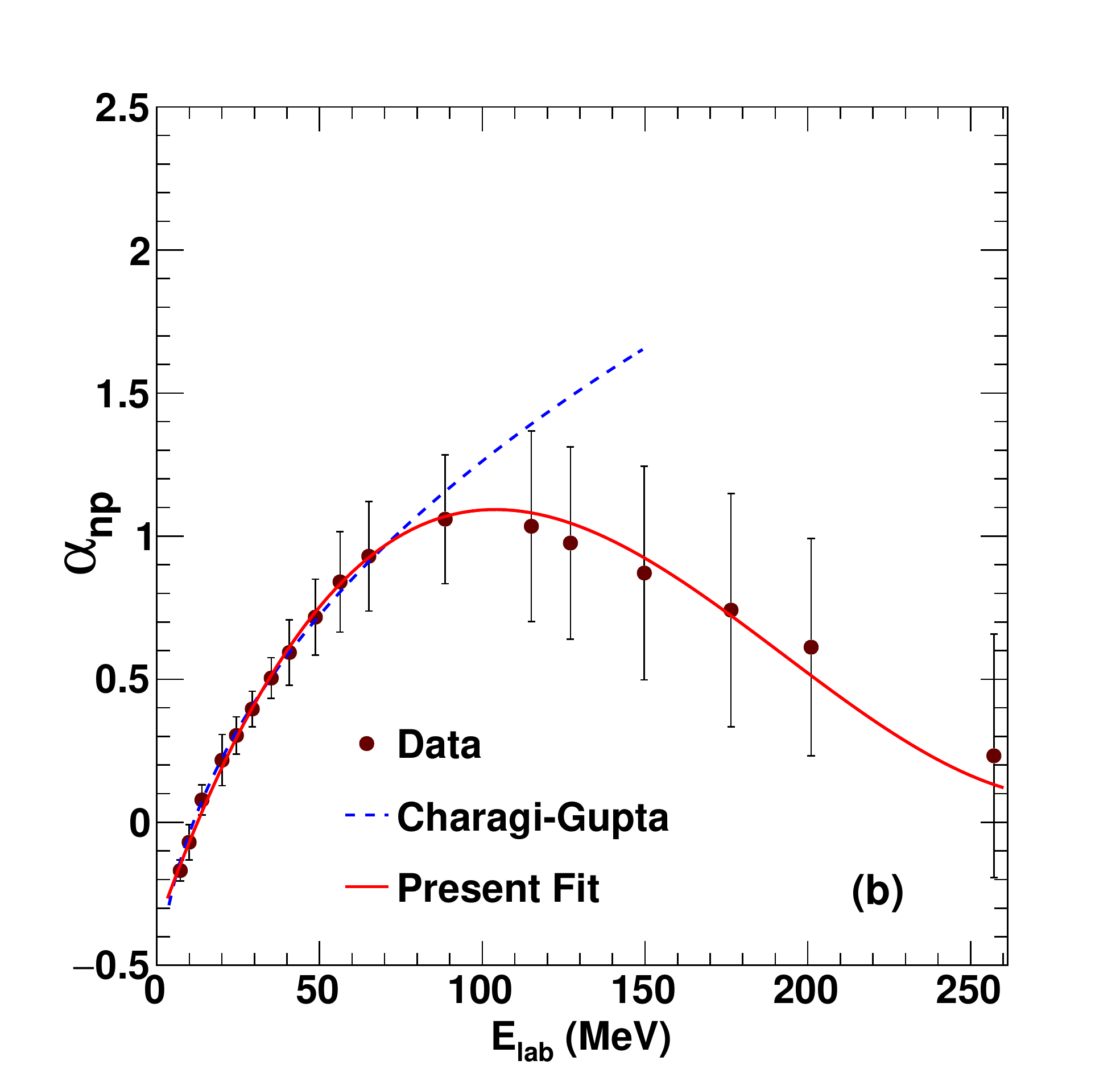}
  \label{alphapp}
  \caption{The ratio of real to imaginary part of NN scattering amplitude 
(a) $\alpha_{pp}$ and (b) $\alpha_{np}$ as a function of lab kinetic energy obtained from 
phase shift analysis
of Ref.~\cite{WGREIN} fitted with the functions given by Eqs.~(\ref{alpp}) and (\ref{alnp})
along with the fit which was given in Ref.~\cite{CHAGUP}.}
  \label{pp_np_phase_shift}
\end{figure}

 Figure~\ref{pp_np_phase_shift} shows the ratio of real to imaginary part of NN scattering 
amplitude as a function of lab kinetic energy
from phase shift analysis of Ref.~\cite{WGREIN} fitted with the functions given 
by Eqs.~(\ref{alpp}) and (\ref{alnp}) along with the fit which was given in Ref.~\cite{CHAGUP}.
The earlier parametrization \cite{CHAGUP} for $\alpha_{pp}$ and $\alpha_{np}$ were good below 60 MeV.

\section{Results and discussions}

  We calculate $\sigma_{R}$ as a function of lab kinetic energy per nucleon $E/A$ 
and $\frac{d\sigma_{el}}{d\sigma_{Ruth}}$ as a function of $\theta_{\rm cm}$ for 
various reaction combinations of light, medium and heavy nuclei and compare with the data.
The reaction cross section data are obtained by optical model analysis of measured
elastic scattering angular distributions. Such analysis is mostly provided by the experimental
group. In case the error on the cross section is not given a 5 \%  error is assumed which is
typically the error obtained in such analysis. The errors on the input parameters, NN cross
sections and density parameters are propagated in the final calculations. The uncertainty
bands also include an 8 \% variation in the nuclear range parameter $r_0$ arround 0.6 fm.

\begin{figure}
\begin{center}
  \includegraphics[width=0.49\textwidth]{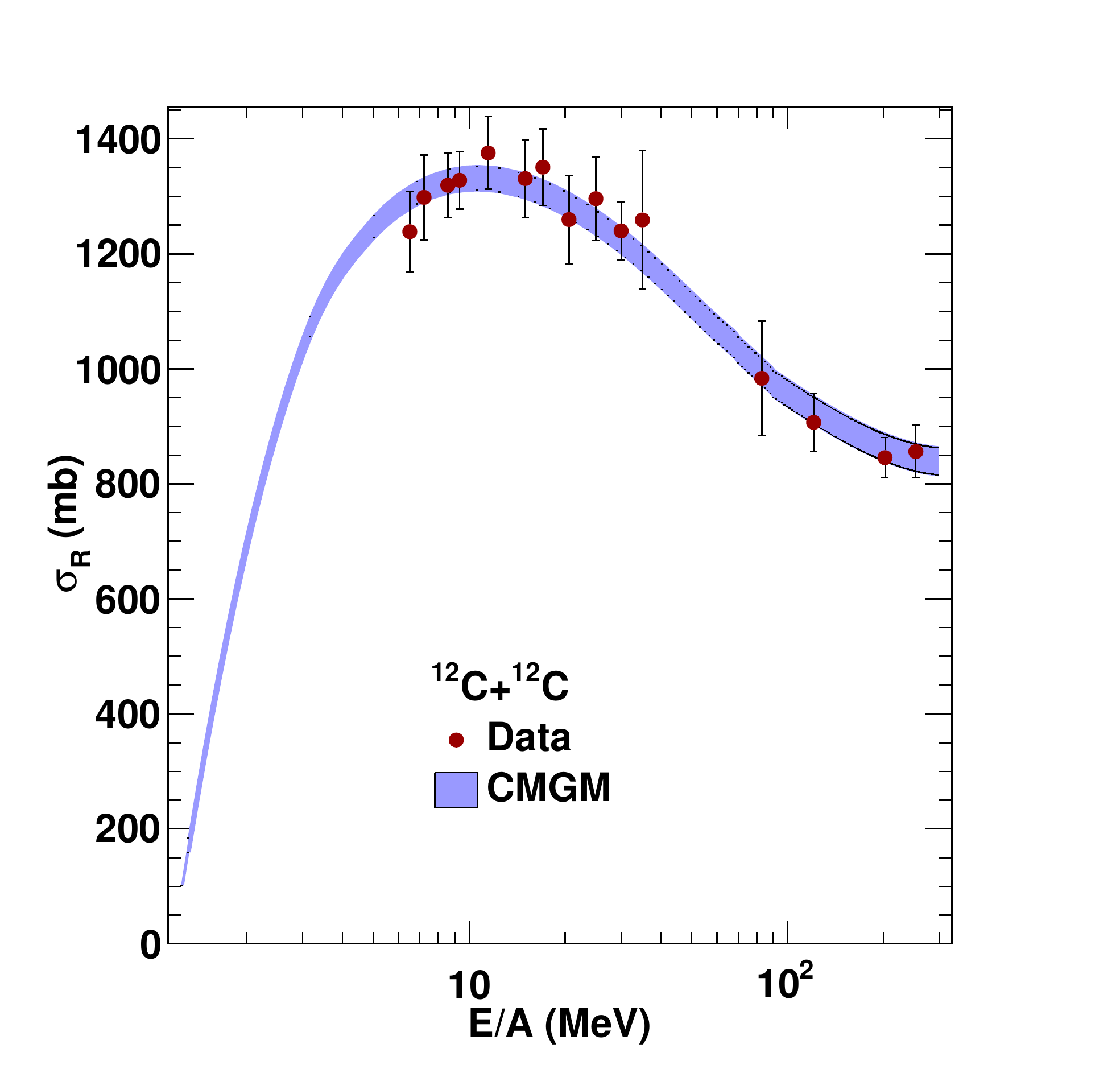}
  \caption{The total reaction cross section as a function of lab kinetic energy per nucleon 
    for $^{12}$C + $^{12}$C system \cite{CCSAHM, Hassanain:2008zz, HISTACHY}.}
  \label{sigmaRfig_C+C}
\end{center}
\end{figure}

\begin{figure}
  \includegraphics[width=0.96\textwidth]{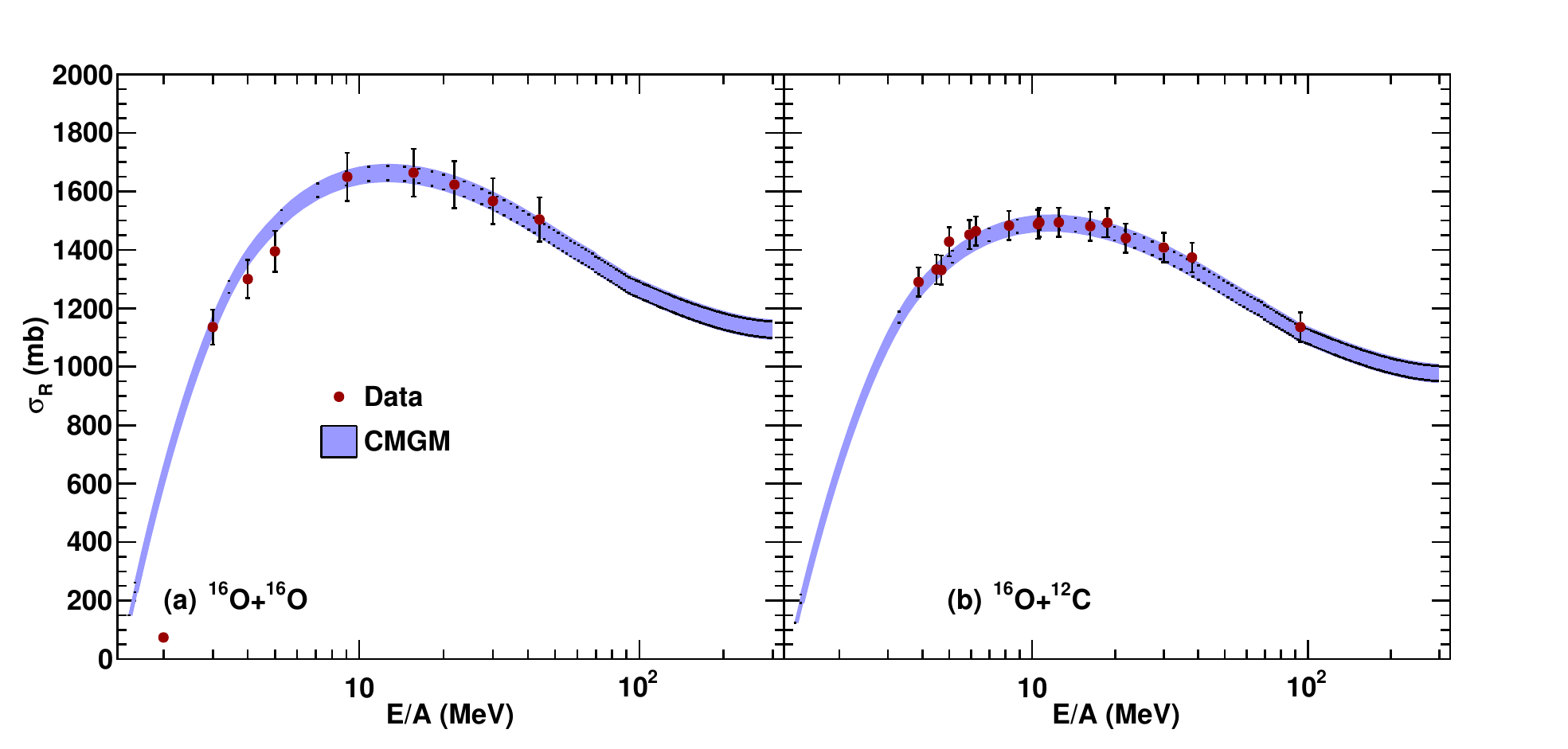}
  \caption{The total reaction cross section as a function of lab kinetic energy per nucleon 
    for (a) $^{16}$O + $^{16}$O system \cite{BASS,PHYREVLET74,HASFARID} and
    (b) $^{16}$O + $^{12}$C system \cite{NICSAT,AZABMAHMOUD,OGLOBIN,SATLOVE,BRANDAN,ROUSSEL1988}.}
  \label{sigmaRfig_O+O}
\end{figure}

  Figure \ref{sigmaRfig_C+C} shows the total reaction cross section as a function of lab kinetic
  energy per nucleon for $^{12}$C + $^{12}$C system \cite{CCSAHM, Hassanain:2008zz, HISTACHY}.
 For this system we have put 5 \% error on the cross section data at $E/A$ = 8.567, 30 and 120.75 MeV.
  Figure \ref{sigmaRfig_O+O} (a) shows the total reaction cross section as a function of 
lab kinetic energy per nucleon for $^{16}$O + $^{16}$O system \cite{BASS,PHYREVLET74,HASFARID} and
the Fig.~\ref{sigmaRfig_O+O} (b) shows the same for
$^{16}$O + $^{12}$C system \cite{NICSAT,AZABMAHMOUD,OGLOBIN,SATLOVE,BRANDAN,ROUSSEL1988}.
 The band is CMGM which includes the uncertainties on the input parameters.
  The model gives very good description of the data for all the light systems except 
at very low energies far below Coulomb barrier for $^{16}$O + $^{16}$O system.

\begin{figure}
  \includegraphics[width=0.96\textwidth]{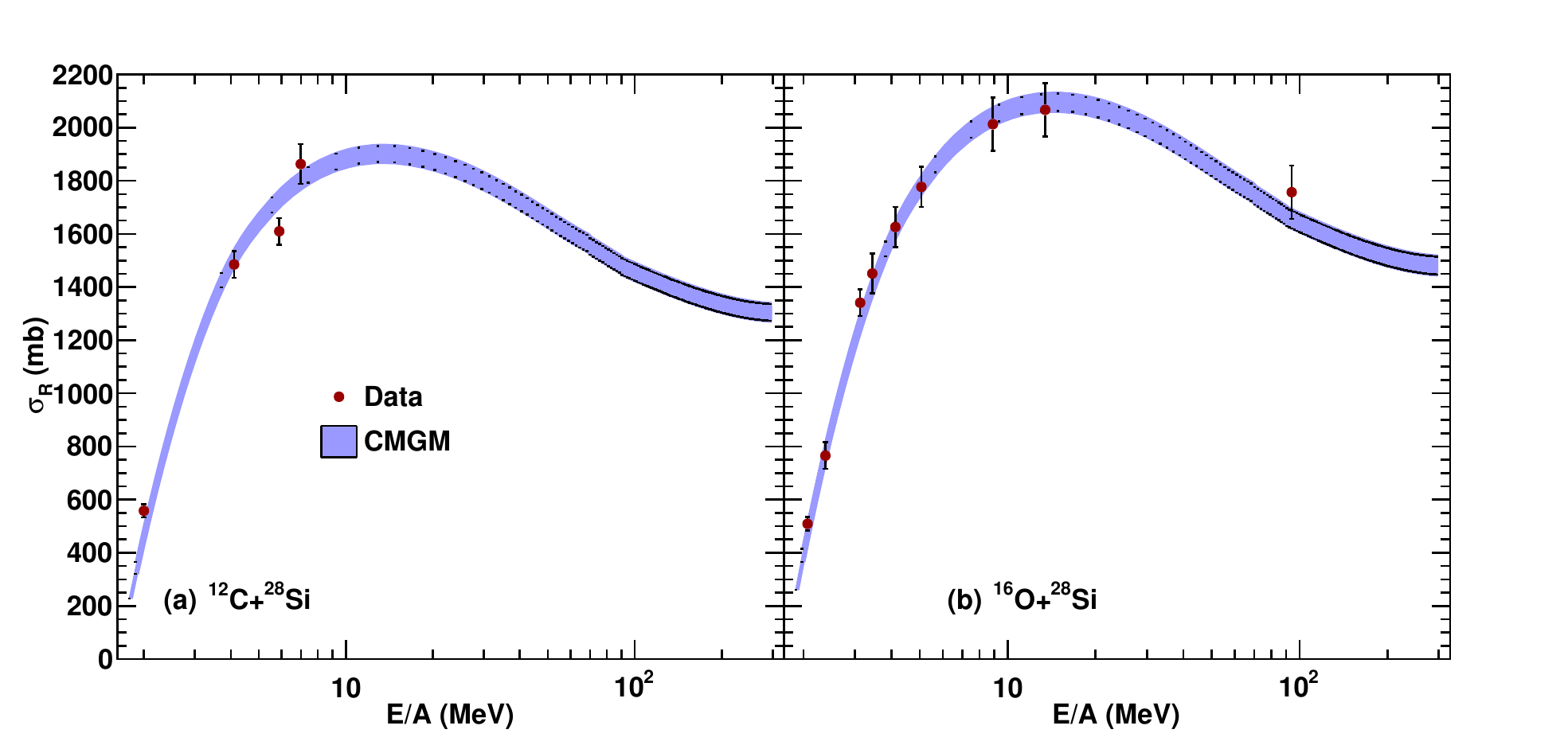}
  \caption{The total reaction cross section as a function of lab kinetic energy per nucleon 
    for (a) $^{12}$C + $^{28}$Si system \cite{AZABMAHMOUD} and
    (b) $^{16}$O + $^{28}$Si system \cite{CRAMER}.}
  \label{sigmaRfig_C+Si}
\end{figure}

\begin{figure}
  \includegraphics[width=0.98\textwidth]{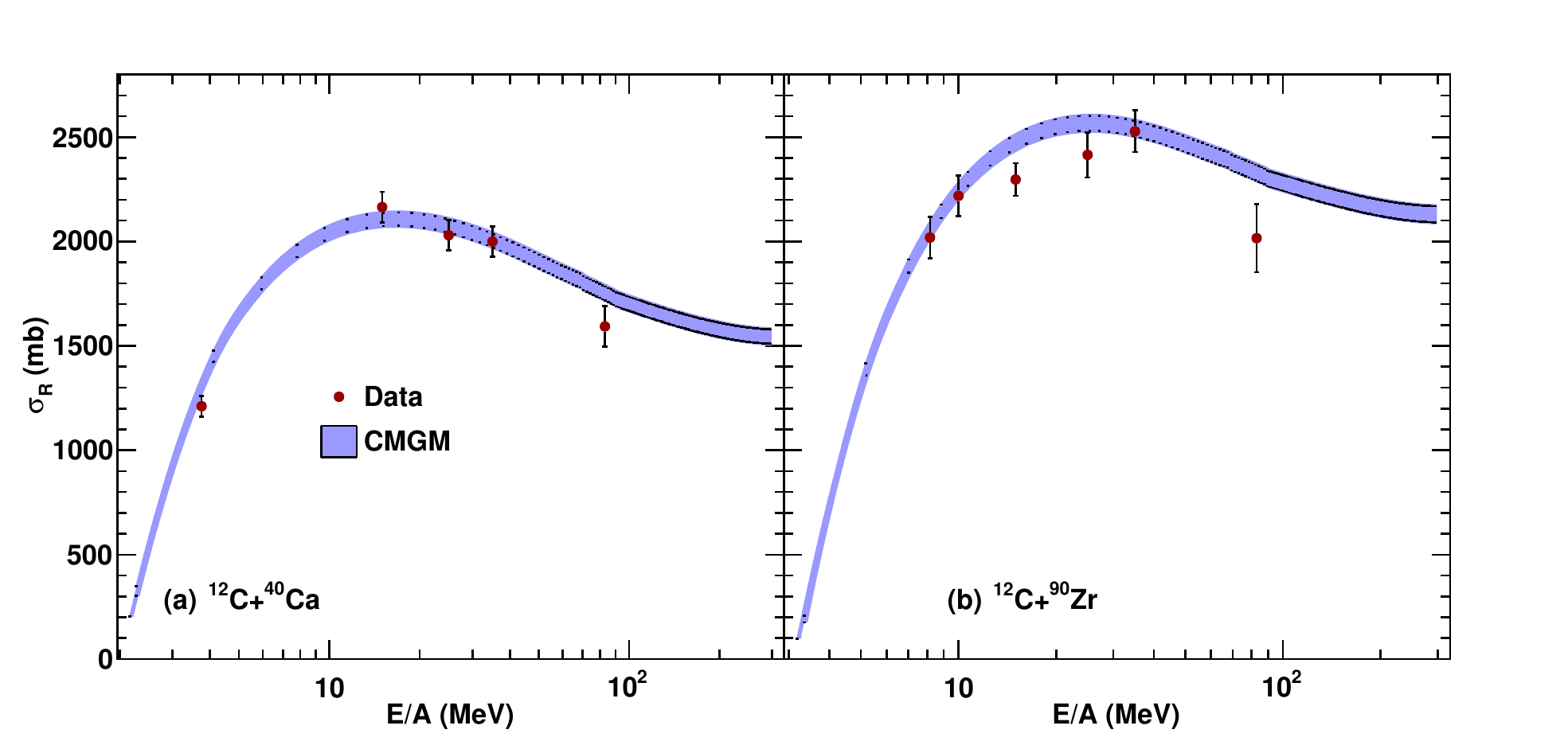}
  \caption{The total reaction cross section as a function of lab kinetic energy per nucleon
    for (a) $^{12}$C + $^{40}$Ca system \cite{CCSAHM,SATLOVE} and
    (b) $^{12}$C + $^{90}$Zr system \cite{CCSAHM,SATLOVE,BRANDAN1997}.}
  \label{sigmaRfig_C+Ca}
\end{figure}

  Figure \ref{sigmaRfig_C+Si} (a)  shows the total reaction cross section as a function of 
lab kinetic energy per nucleon for $^{12}$C + $^{28}$Si system \cite{AZABMAHMOUD} and the 
Fig.~\ref{sigmaRfig_C+Si} (b) shows the same for $^{16}$O + $^{28}$Si system \cite{CRAMER}.
  For all the data shown in figures \ref{sigmaRfig_O+O} and \ref{sigmaRfig_C+Si}
we have put 5 \% error on the reaction cross section.
 Figure \ref{sigmaRfig_C+Ca} (a) shows the total reaction cross section as a function of 
lab kinetic energy per nucleon for $^{12}$C + $^{40}$Ca system \cite{CCSAHM,SATLOVE} and the 
Fig.~\ref{sigmaRfig_C+Ca} (b) shows the same for
$^{12}$C + $^{90}$Zr system \cite{CCSAHM,SATLOVE,BRANDAN1997}.
The band is CMGM which includes the uncertainties on the input parameters.
  The model gives good description of the data for medium mass systems except for 
the $^{12}$C + $^{90}$Zr system. 
For $^{12}$C + $^{40}$Ca system at energy $E/A$ = 3.75 MeV we have put a 5 \% error on the
 reaction cross section. Similar error was put in the cross section for the system 
$^{12}$C + $^{90}$Zr at energies $E/A$ = 8.166 and 35 MeV.

\begin{figure}
  \includegraphics[width=0.98\textwidth]{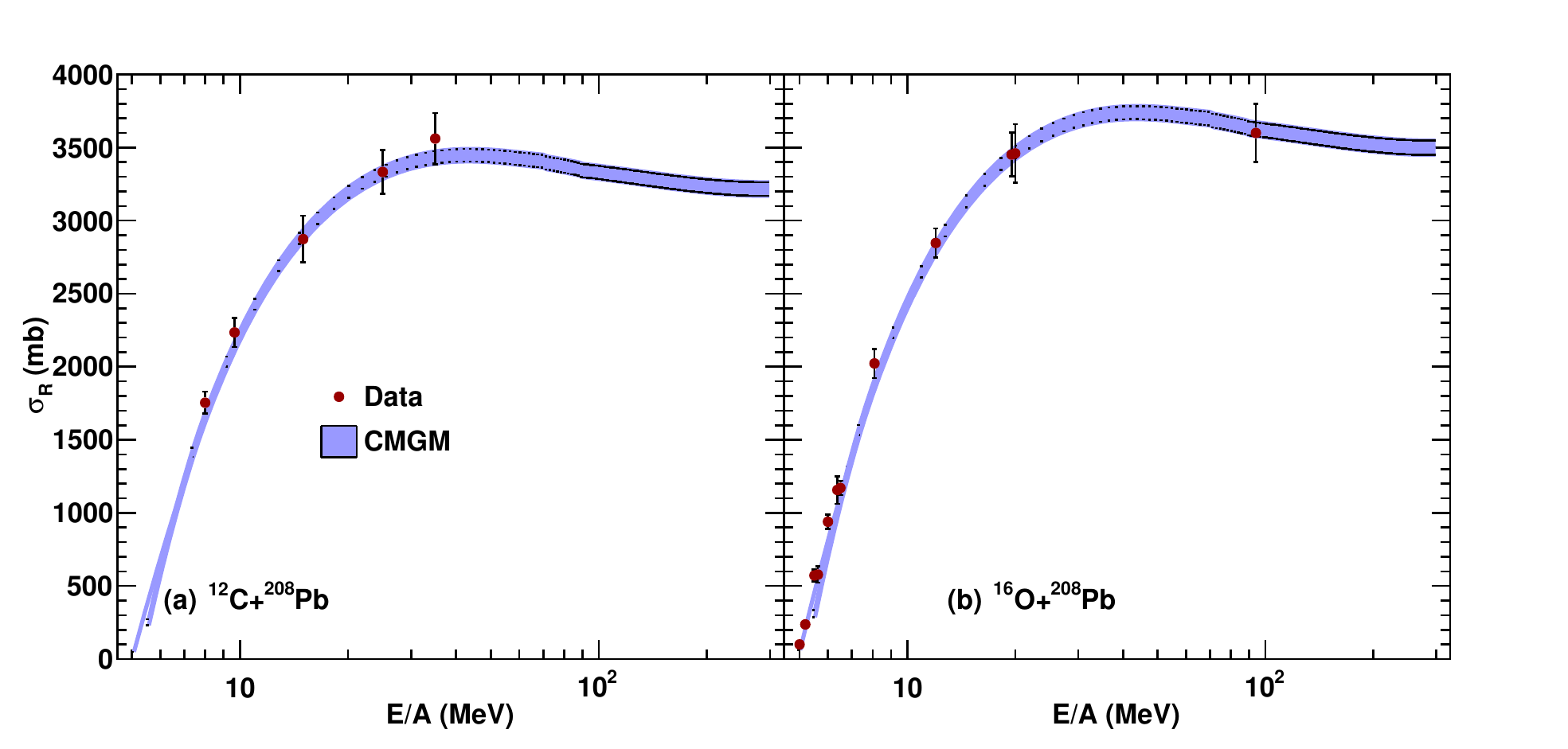}
  \caption{The total reaction cross section as a function of lab kinetic energy per nucleon
    for (a) $^{12}$C + $^{208}$Pb system \cite{CCSAHM,BRANDAN1997,PHYLETB102}
    and (b) $^{16}$O + $^{208}$Pb system \cite{VAZBACH,BECHETI,BALL,CHARAGI1995,Olmer1978,TMMEHIP,ROUSSEL1988}.}
  \label{sigmaRfig_C+Pb}
\end{figure}

  Figure \ref{sigmaRfig_C+Pb} (a) shows the total reaction cross section as a function of 
lab kinetic energy per nucleon for $^{12}$C + $^{208}$Pb system \cite{CCSAHM,BRANDAN1997,PHYLETB102}.
  Figure \ref{sigmaRfig_C+Pb} (b) shows the same for 
$^{16}$O + $^{208}$Pb system \cite{VAZBACH,BECHETI,BALL,CHARAGI1995,Olmer1978,TMMEHIP}.
The band is CMGM which includes the uncertainties on the input parameters.
  For $^{12}$C + $^{208}$Pb system at energies $E/A$ = 8 and 9.66 MeV we have put a 5 \% error on the
reaction cross section. Similar error was put in the cross section for the system 
$^{16}$O + $^{208}$Pb at energies $E/A$ = 6.0, 8.093, 12.0 and 19.54 MeV.

  The model gives very good description of the data for both the heavy systems considered here.
We can conclude that the reaction cross section calculations from the model 
are very reliable and and thus the model can be used to predict the reaction cross section 
for unknown systems. It can also be used to obtain radii of nuclei from measured 
reaction cross section.

\clearpage

\begin{figure}
  \includegraphics[width=0.98\textwidth]{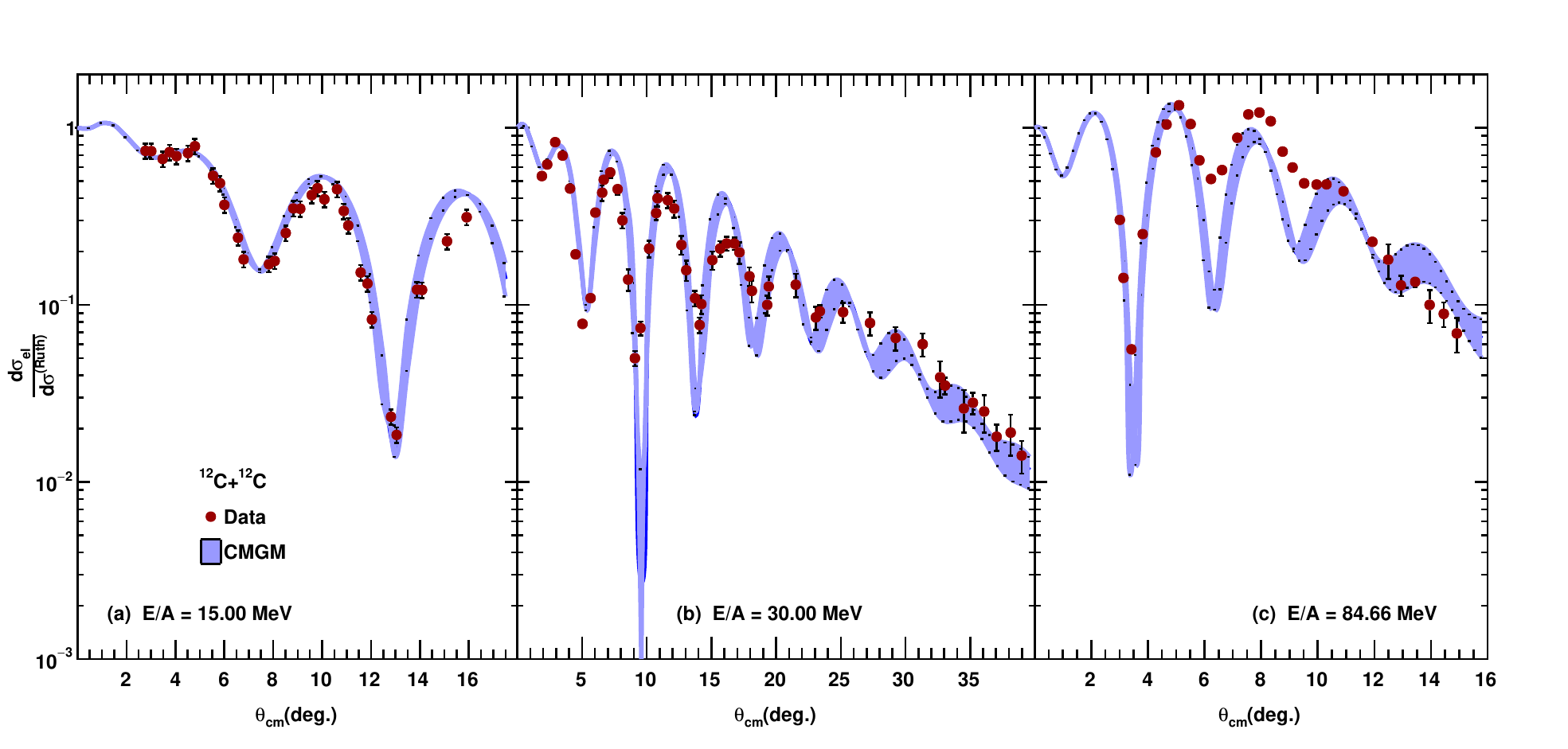}
  \caption{The measured ratio of elastic scattering cross section to the 
Rutherford cross section as a function of scattering angle for $^{12}$C + $^{12}$C system
at (a) $E/A$ = 15 MeV \cite{PRC491652}, (b) $E/A$ = 30 MeV 
\cite{PHYLETB102,PHYREVC26,NUCPHYA424} and (c) $E/A$ = 84.66 MeV \cite{PHYLETB102,PHYREVC26}
along with the CMGM calculations.}
\label{figelat_C+C}
\end{figure}
\begin{figure}
  \includegraphics[width=0.98\textwidth]{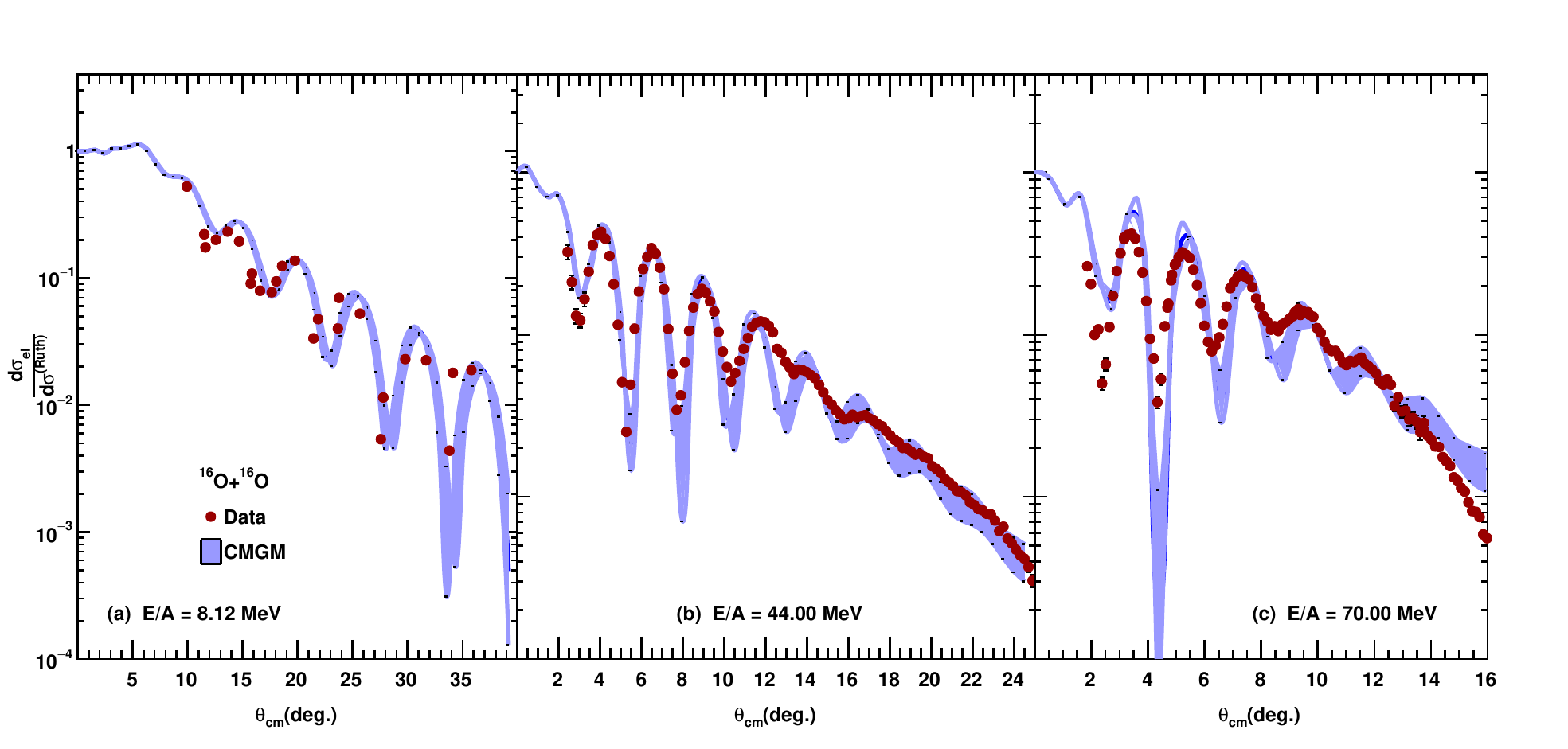}
  \caption{The measured ratio of elastic scattering cross section to the 
Rutherford cross section as a function of scattering angle for $^{16}$O + $^{16}$O system
at (a) $E/A$ =  8.12 MeV \cite{NUCPHYA456}
(b) $E/A$ =  44 MeV \cite{NUOCIMA111,PHYLETB365} and (c) $E/A$ = 70 MeV \cite{NUOCIMA111}
along with the CMGM calculations.}
  \label{figelat_O+O}
\end{figure}

  Figures \ref{figelat_C+C} shows the measured ratio of elastic scattering cross section to the 
Rutherford cross section as a function of scattering angle for $^{12}$C + $^{12}$C system
at three energies given in (a) $E/A$ = 15 MeV \cite{PRC491652}, (b) $E/A$ = 30 MeV 
\cite{PHYLETB102,PHYREVC26,NUCPHYA424} and (c) $E/A$ = 84.66 MeV \cite{PHYLETB102,PHYREVC26}
along with the CMGM calculations shown by bands.
  Figure \ref{figelat_O+O} shows the measured ratio of elastic scattering cross section to the 
Rutherford cross section as a function of scattering angle for $^{16}$O + $^{16}$O system
for three energies given in (a) $E/A$ =  8.12 MeV \cite{NUCPHYA456}
(b) $E/A$ =  44 MeV \cite{NUOCIMA111,PHYLETB365} and (c) $E/A$ = 70 MeV \cite{NUOCIMA111}
along with the CMGM calculations shown by bands.


 The parameter $\bar \alpha_{NN}$ is obtained by fitting the experimental data on angular
distribution which is given in the Table~\ref{TablealphaNN} along with the values
obtained using parametrizations given by Eqs.~(\ref{alpp}) and (\ref{alnp}).
  The model produces the measured diffractive oscillations but the oscillation magnitudes in
  the data in the light systems are more pronounced as compared to the oscillations observed
  in the data specially at higher energies and higher angles i.e. at the higher momentum transfer.
The parameter $\bar \alpha_{NN}$ does not control oscillation magnitude, but affects
the slope (inclination) of $d\sigma/d\sigma_{Ruth}$ as a function of $\theta_{cm}$.

\begin{figure}
  \includegraphics[width=0.98\textwidth]{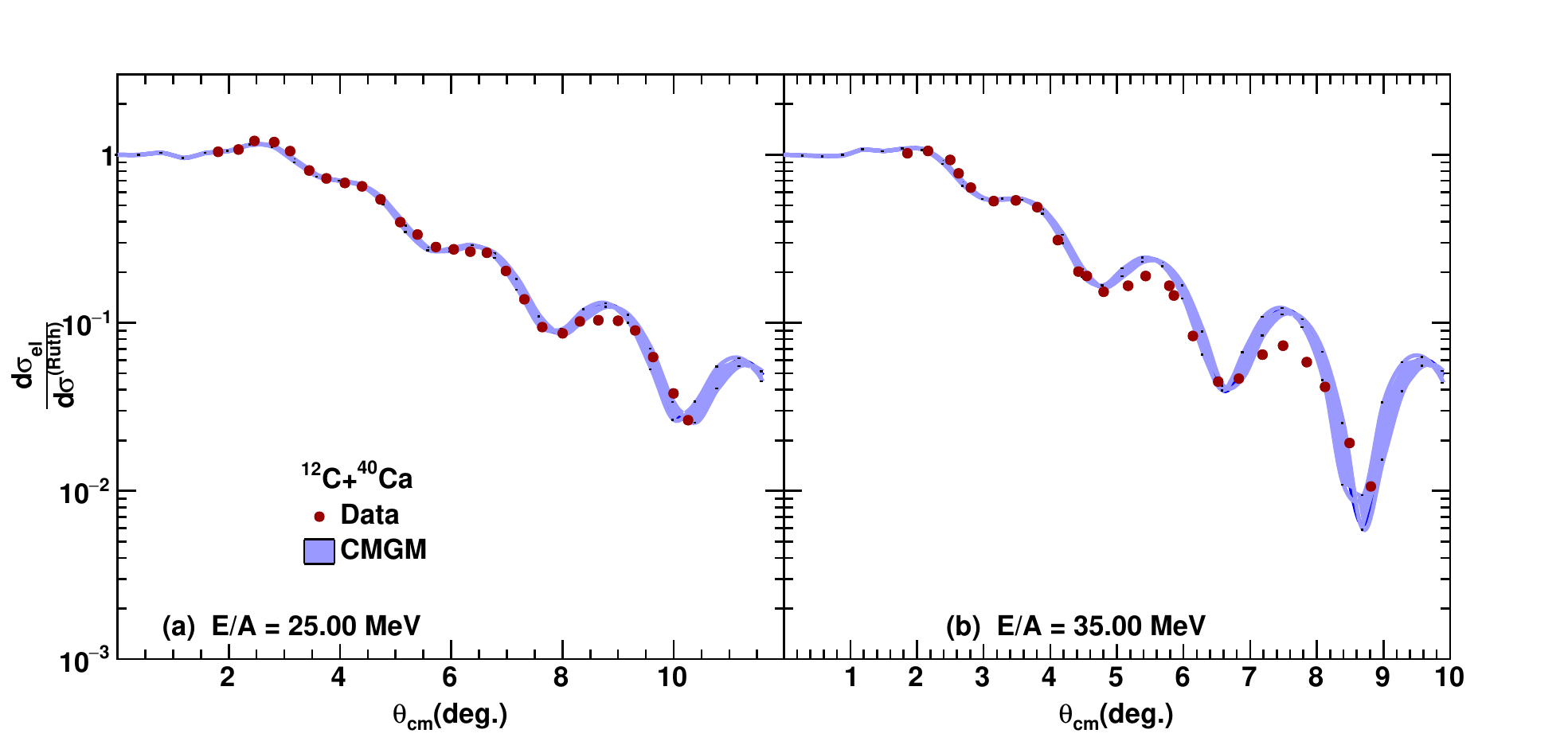}
  \caption{The measured ratio of elastic scattering cross section to the 
Rutherford cross section as a function of scattering angle for $^{12}$C + $^{40}$Ca system
\cite{CCSAHM} at (a) $E/A$ = 25 MeV and (b) $E/A$ = 35 MeV 
along with the CMGM calculations shown by bands.}
\label{figelat_C+Ca}
\end{figure}

Figure \ref{figelat_C+Ca} shows the measured ratio of elastic scattering cross section to the 
Rutherford cross section as a function of scattering angle for $^{12}$C + $^{40}$Ca system
\cite{CCSAHM} for two energies given in (a) $E/A$ = 25 MeV and (b) $E/A$ = 35 MeV 
along with the CMGM calculations shown by bands.

\begin{figure}
  \includegraphics[width=0.98\textwidth]{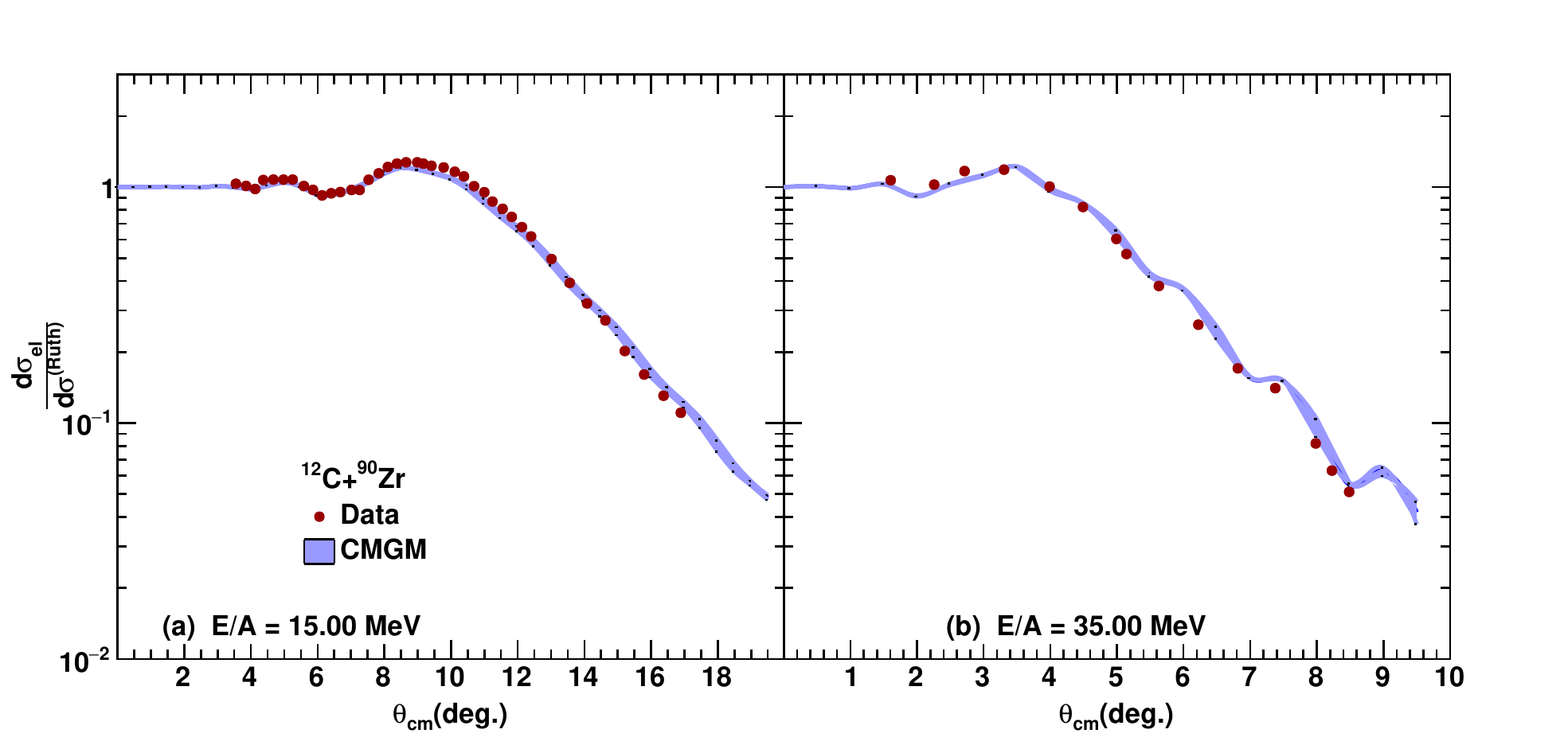}
  \caption{The measured ratio of elastic scattering cross section to the 
Rutherford cross section as a function of scattering angle for $^{12}$C + $^{90}$Zr system
\cite{CCSAHM} at (a) $E/A$ = 15 MeV  and (b) $E/A$ = 35 MeV
along with the CMGM calculations shown by bands.}
\label{figelat_C+Zr}
\end{figure}

Figure \ref{figelat_C+Zr} shows the measured ratio of elastic scattering cross section to the 
Rutherford cross section as a function of scattering angle for $^{12}$C + $^{90}$Zr system
\cite{CCSAHM} for two energies given in (a) $E/A$ = 15 MeV  and (b) $E/A$ = 35 MeV
along with the CMGM calculations shown by bands.

\begin{figure}
  \includegraphics[width=0.98\textwidth]{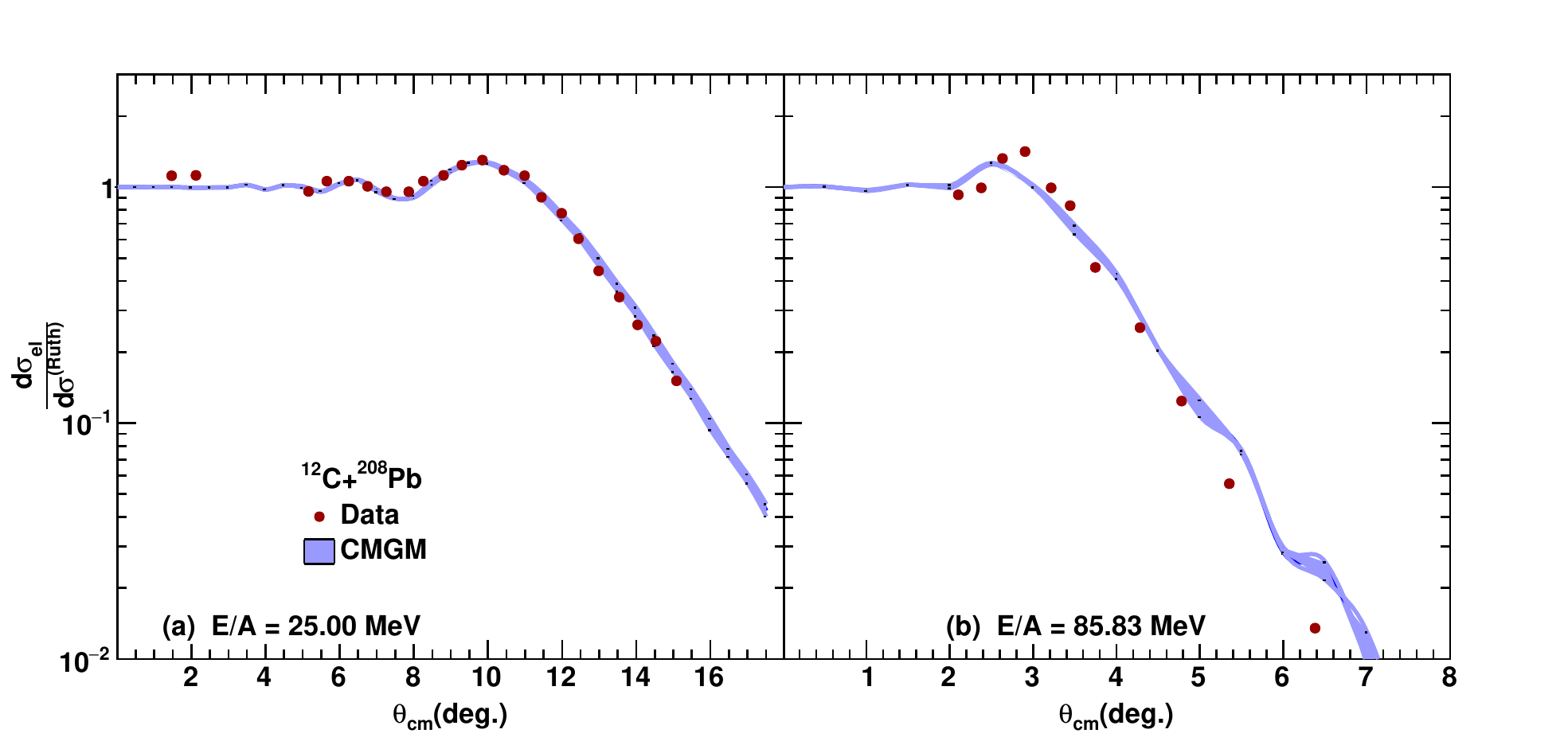}
  \caption{The measured ratio of elastic scattering cross section to the 
Rutherford cross section as a function of scattering angle for $^{12}$C + $^{208}$Pb system
\cite{CCSAHM,ALVAREZ} at (a) $E/A$ = 25 MeV  and (b) $E/A$ = 85.83 MeV
along with the CMGM calculations shown by bands.}
\label{figelat_C+Pb}
\end{figure}

Figure \ref{figelat_C+Pb} shows the measured ratio of elastic scattering cross section to the 
Rutherford cross section as a function of scattering angle for $^{12}$C + $^{208}$Pb system
\cite{CCSAHM,ALVAREZ} for two energies given in (a) $E/A$ = 25 MeV  and (b) $E/A$ = 85.83 MeV
along with the CMGM calculations shown by bands.

\begin{figure}
  \includegraphics[width=0.98\textwidth]{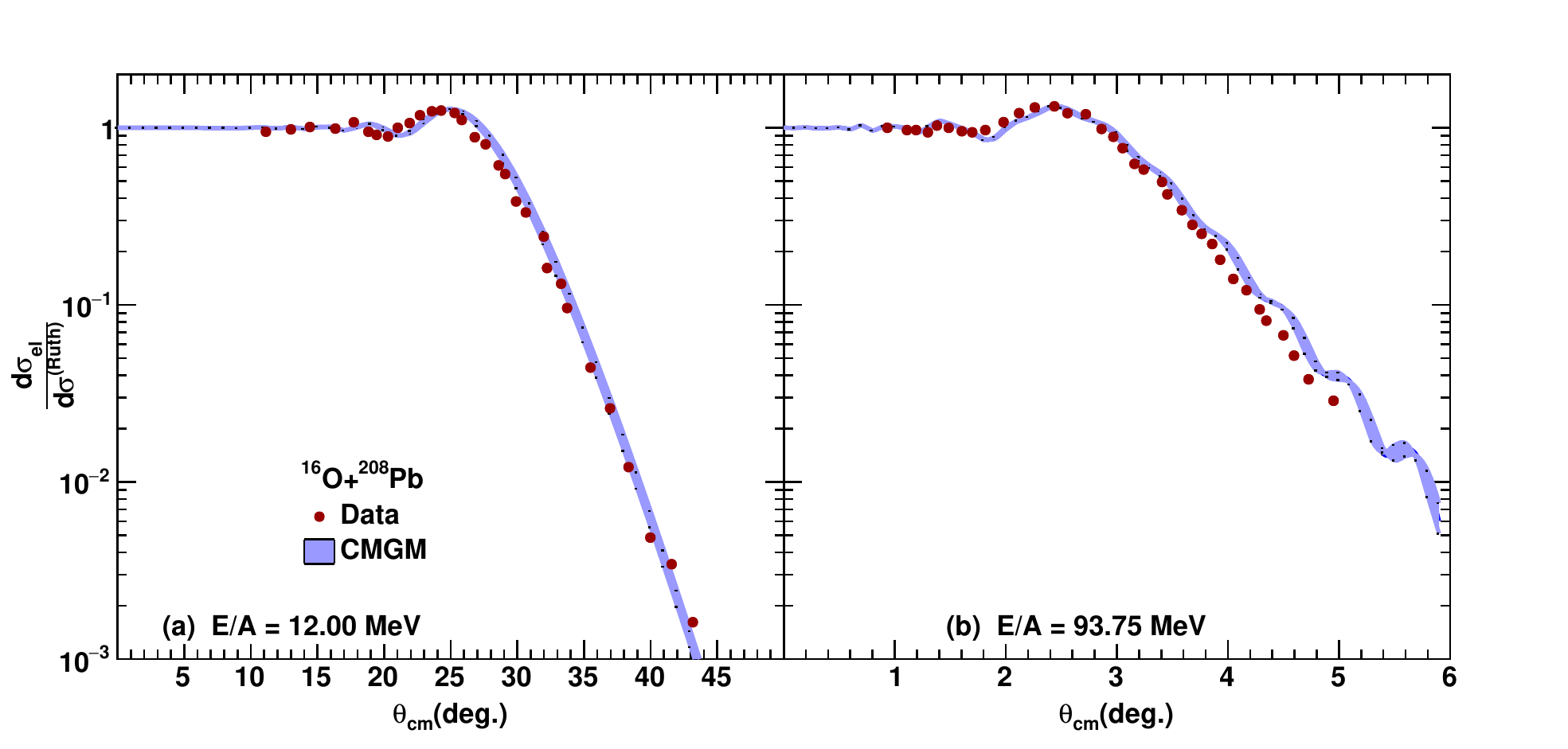}
  \caption{The measured ratio of elastic scattering cross section to the 
Rutherford cross section as a function of scattering angle for $^{16}$O + $^{208}$Pb system
\cite{CHARAGI1995,ROUSSEL1988} at (a) $E/A$ = 12 MeV  and (b) $E/A$ = 93.75 MeV
along with the CMGM calculations shown by bands.}
\label{figelat_O+Pb}
\end{figure}

Figure \ref{figelat_O+Pb} shows the measured ratio of elastic scattering cross section to the 
Rutherford cross section as a function of scattering angle for $^{16}$O + $^{208}$Pb system
\cite{CHARAGI1995,ROUSSEL1988} for two energies given in (a) $E/A$ = 12 MeV  and (b) $E/A$ = 93.75 MeV
along with the CMGM calculations shown by bands.

  The fitted values of the parameter $\bar \alpha_{NN}$ for all the systems are given in the 
Table~\ref{TablealphaNN}. For heavy ion systems this parameter does not follow the same 
energy dependence shown by NN scattering and approaches towards one for all the systems.
  The model produces the measured elastic scattering angular distributions for medium and 
heavy systems at low energies but, at higher energies the model produces diffractive 
oscillations of larger magnitude as compared to the data.

  The large oscillations in the model at high momentum transfer may be the consequence of
assuming a semiclassical picture of scattering in terms of impact parameter and distance
of closest approach.
 The Glauber optical potential in terms of densities of the two nuclei and NN amplitude
does not always reproduce the features of data at large scattering angles. 
 An improved fitting should be obtained by having more free parameters in the
nuclear potential/interactions shown in the studies
in Ref.~\cite{DTKhao:2000npa} or in Ref.~\cite{Ahmad:2004alvi}.

\clearpage 




\clearpage

\begin{table}
\caption[]{Calculated (from parametrization) and fitted values of $\bar \alpha_{NN}$.}
\label{TablealphaNN}
\begin{tabular}{l l l l}
\hline
\hline
    System          &      $E/A$ (MeV)   &    $\bar \alpha_{\rm NN(calc)}$ &   $\bar \alpha_{\rm NN(fitted)}$ \\
\hline
$^{12}$C + $^{12}$C   &       15.00        &        0.255           &          0.9        \\         
                    &       30.00        &        0.595           &          0.78       \\       
                    &       84.66        &        1.25            &          0.9        \\
$^{16}$O + $^{16}$O   &       8.12         &        0.1             &          0.80       \\     
                    &       44.00        &        0.85            &          0.94       \\     
                    &       70.00        &        1.165           &          1.02       \\  
$^{12}$C + $^{40}$Ca  &       25.00        &        0.488           &          0.8        \\  
                    &       35.00        &        0.694           &          1.0        \\        
$^{12}$C + $^{90}$Zr  &       15.00        &        0.255           &          0.7        \\     
                    &       35.00        &        0.694           &          1.0        \\
$^{12}$C + $^{208}$Pb &       25.00        &         0.48           &          1.0         \\      
                    &       85.83        &        1.25           &          1.254      \\     
$^{16}$O + $^{208}$Pb &       12.00        &         0.181          &          1.0        \\  
                    &       93.75        &        1.272           &          1.5        \\
\hline
\hline
\end{tabular}
\end{table}

\section{Conclusions}
  We calculate the heavy ion reaction cross sections and elastic scattering 
angular distributions at low and intermediate energies using Glauber model and 
test the calculations with the available measured data. 
  We present new parametrization for the data of total cross sections
  as well as for the ratio of real to imaginary parts of the scattering amplitudes
  in case of pp and np collisions.
 The model works at low energies down to Coulomb barrier with very simple 
modifications.
 The reaction cross section calculations from the model 
are very reliable and thus the model can be used to predict the reaction cross section 
for unknown systems. It can also be used to obtain radii of nuclei from measured 
reaction cross sections in low and intermediate energy range.

 The model describes the measured elastic scattering angular distributions
having diffractive oscillations but the oscillation magnitude in the data
is more pronounced as compared to the oscillations observed in the data 
specially at higher energies and higher angles i.e. at the higher momentum transfer.
  For heavy ion systems, the parameter $\bar \alpha_{NN}$ does not follow the same 
energy dependence shown by NN scattering and approaches towards one for all the systems.


\section*{References}


\begin{thebibliography}{00}


\bibitem{GLAUBER} R. J. Glauber, {\it Lectures on theoretical Physics}, Vol. I, 
                 Inter-Science, New York (1959).

\bibitem{KAROL} P. J. Karol, Phys. Rev. C{\bf 11} (1975) 1203.

\bibitem{WONG} C.Y. Wong, {\it Introduction to High Energy Heavy Ion Collisions}, World Scientific, Singapore (1994).

\bibitem{SHUKLAarxiv2001} P. Shukla, Preprint: nucl-th/0112039 (2001).

\bibitem{d'Enterria:2003qs} 
  D.~G.~d'Enterria,
  nucl-ex/0302016.

\bibitem{CMGM}  A. Vitturi and F. Zardi, Phys. Rev. C{\bf 36} (1987) 1404;
     S.M. Lenzi, A. Vitturi, F. Zardi, Phys. Rev. C{\bf 40} (1989) 2114.

\bibitem{CHAGUP} S. K. Charagi and S. K. Gupta, Phys. Rev. C{\bf 41} (1990) 1610; 
            Phys. Rev. C{\bf 46} (1992) 1982.

\bibitem{SHUKLA1995} S. K. Gupta and P. Shukla, Phys. Rev. C{\bf 52} (1995) 3212.


\bibitem{WARNER} R. E. Warner, M. H. McKinnon, and H. Thirumurthy and A. Nadasen, Phys. Rev. C{\bf 59} (1999) 1215.

\bibitem{AHMAD} I. Ahmad, M. A. Abdulmomen, and M. S. Al-Enazi, Phys. Rev. C{\bf 65} (2002) 054607.

\bibitem{CAI}  X.~Z.~Cai, J.~Feng, W.~Q.~Shen, Y.~G.~Ma, J.~S.~Wang and W.~Ye,
  Phys.\ Rev.\ C {\bf 58} (1998) 572.

\bibitem{PNU} A. de Vismes, P. Roussel-Chomaz, and F. Carstoiu, Phys. Rev. C{\bf 62} (2000) 064612.

\bibitem{SHUKLAprc2003} P. Shukla, Phy. Rev. C{\bf 67} (2003) 054607.


\bibitem{Alkhazovi:2011ty}  G.~D.~Alkhazovi, Y.~Shabelski and I.~S.~Novikov,
  Int.\ J.\ Mod.\ Phys.\ E {\bf 20} (2011) 583
  [arXiv:1101.4717 [nucl-th]].

\bibitem{TANIHATA} I. Tanihata et al., Phys. Rev. Lett. {\bf 55} (1985) 2676; 
   I. Tanihata, Nucl. Phys. A{\bf 488} (1988) 113c; I. Tanihata et. al, Phys. Lett. B{\bf 289} (1992) 261.


\bibitem{Horiuchi:2006ga} 
  W.~Horiuchi, Y.~Suzuki, B.~Abu-Ibrahim and A.~Kohama,
  Phys.\ Rev.\ C {\bf 75} (2007) 044607. 
   Erratum: [Phys.\ Rev.\ C {\bf 76} (2007) 039903].
  [nucl-th/0612029].

\bibitem{PATRA2009} S. K. Patra, R. N. Panda, P. Arumugam and R. K. Gupta, 
Phys.\ Rev.\ C{\bf 80} (2009) 064602.

\bibitem{Sharma:2013nba} 
  M.~K.~Sharma and S.~K.~Patra,
  Phys.\ Rev.\ C{\bf 87} (2013) 044606.

\bibitem{Chauhan:2014uoa} 
  D.~Chauhan, Z.~A.~Khan and A.~A.~Usmani,
  Phys.\ Rev.\ C {\bf 90} (2014) 024603.


\bibitem{Hassan:2009dc} 
  M.~Y.~M.~Hassan, M.~Y.~H.~Farag, A.~Y.~Abul-Magd and T.~E.~I.~Nassar,
  Phys.\ Scripta {\bf 78} (2008) 045202  [arXiv:0902.0453 [nucl-th]].


\bibitem{Christley:1999jat} 
  J.~A.~Christley and J.~A.~Tostevin, 
 Phys.\ Rev.\ C {\bf 59} (1999) 2309.
 


\bibitem{Lukyanov:2015aya} 
  V.~K.~Lukyanov, D.~N.~Kadrev, E.~V.~Zemlyanaya, K.~Spasova, 
K.~V.~Lukyanov, A.~N.~Antonov and M.~K.~Gaidarov,
  Phys.\ Rev.\ C {\bf 91} (2015) 034606.
  [arXiv: 1502.06425 [nucl-th]].




\bibitem{DTKhao:2000npa} 
Dao~T.~Khoa, ~W. von Oertzen, ~H.~G. Bohlena and F. Nuofferc, 
Nucl.\ Phys.\ A {\bf 672} (2000) 387.

\bibitem{ElGogary:2000cv} 
  M.~M.~H.~El-Gogary, A.~S.~Shalaby, M.~Y.~M.~Hassan and A.~M.~Hegazy,
  Phys.\ Rev.\ C {\bf 61} (2000) 044604.

\bibitem{ElGogary:1998gu} 
  M.~M.~H.~El-Gogary, A.~S.~Shalaby and M.~Y.~M.~Hassan,
  Phys.\ Rev.\ C {\bf 58} (1998) 3513.

\bibitem{Sammarruca:2011nc} 
  F.~Sammarruca and L.~White,
  Phys.\ Rev.\ C {\bf 83} (2011) 064602.
  [arXiv: 1105.5666 [nucl-th]].  

\bibitem{Ahmad:2004alvi} 
  I.~AHMAD and M.~A.~ALVI, 
   Int.\ J.\ Mod.\ Phys.\ E {\bf 13} (2000) 1225.

\bibitem{Gibbs:2012yd} 
  W.~R.~Gibbs and J.~P.~Dedonder,
  Phys.\ Rev.\ C {\bf 86} (2012) 024604.
  [arXiv: 1203.0019 [nucl-th]].

\bibitem{Bertulani:2010kk} 
  C.~A.~Bertulani and C.~De Conti,
  Phys.\ Rev.\ C {\bf 81} (2010) 064603.
  [arXiv:1004.2096 [nucl-th]].
  
\bibitem{JAGER1974} C. W. de Jager, H. de Vries and C. de Vries, Atom. Nucl. Data Tables {\bf 14} (1974) 479.
  
\bibitem{JAGER1987} H. De Vries, C. W. de Jager, and C. de Vries, Atom. Nucl. Data Tables {\bf 36} (1987) 495.

\bibitem{Franco1987} V. Franco and A. Tekou, Phys. Rev. C{\bf 16} (1987) 658.


\bibitem{PDG} Particle Data Group, (http://pdg.lbl.gov/2014/hadronic-xsections/). 

\bibitem{WGREIN} W. Grein, Nucl. Phys. B{\bf 131} (1977) 255.

\bibitem{CCSAHM} C. C. Sahm et. al, Phys. Rev. C{\bf 34} (1986) 2165.

\bibitem{Hassanain:2008zz} 
  M.~A.~Hassanain, A.~A.~Ibraheem and M.~E.~Farid,
  Phys.\ Rev.\ C {\bf 77} (2008) 034601.

\bibitem{HISTACHY} J. Y. Histachy et. al, Nucl. Phys. A{\bf 490} (1998) 441.

\bibitem{BASS} R. Bass, {\it Nuclear reactions with heavy ions}, (Springer-Verlag, NY), 1980.

\bibitem{PHYREVLET74} Dao T. Khoa, W. von Oertzen, H. G. Bohlen, G. Bartnitzky, H. Clement, Y. Sugiyama, 
B. Gebauer, A. N. Ostrowski, Th. Wilpert, M. Wilpert, and C. Langner, Phys. Rev. Lett. {\bf 74} (1995) 34.
  
  
\bibitem{HASFARID} M. A. Hassanain, A. A. Ibraheem, S. M. M. Al Sebiey, S. R. Mokhtar, M. A. Zaki, 
     Z. M. M. Mahmoud, K. O. Behairy and M. E. Farid, Phys. Rev. C{\bf 87} (2013) 064606.

\bibitem{NICSAT} M. P. Nicoli, F. Haas, R. M. Freeman, S. Szilner, Z. Basrak, A. Morsad, G. R. Satchler and M. E. Brandan, 
Phys. Rev. C{\bf 61} (2000) 034609. 

\bibitem{AZABMAHMOUD} M. El-Azab Farid, Z. M. M. Mahmoud and G.S. Hassan, Nucl. Phys. A{\bf 691} (2001) 671.

\bibitem{OGLOBIN} A. A. Ogloblin et. al, Phys. Rev. C{\bf 62} (2000) 044601.

\bibitem{SATLOVE} G. R. Satchler and W.G. Love, Phys. Rep. {\bf 55} (1979) 183.

\bibitem{BRANDAN} M. E. Brandan et. al, Nucl. Phys. A{\bf 688} (2001) 659.

\bibitem{ROUSSEL1988} P. Roussel-Chomaz et. al, Nucl. Phys. A{\bf 477} (1988) 345.

\bibitem{CRAMER} J.G. Cramer, R.M. Devries, D.A. Goldberg, M.A. Zisman
             and C.F. Maguire, Phys. Rev. C{\bf 14} (1976) 2158.

\bibitem{BRANDAN1997} M. E. Brandan, H. Chehime and K. W. McVoy, Phys. Rev. C{\bf 55}
  (1997) 1353. 

             
\bibitem{PHYLETB102} M. Buenerd et. al, Phys. Lett. B{\bf 102} (1981) 242.
             
\bibitem{VAZBACH} Louis C. Vaz, John M. Alexander, E. H. Auerbach, Phys. Rev. C{\bf 18} (1978) 820.

  
\bibitem{BECHETI} F. D. Becheti, Phys. Rev. C{\bf 6} (1972) 2215.
  
\bibitem{BALL} J. B. Ball, C. B. Fulmer, E. E. Gross, M. L. Halbert, D. C. Hensley, C. A. Ludemann, 
  M. J. Saltmarsh and G. R. Satchler, Nucl. Phys. A{\bf 252} (1975) 208.

\bibitem{CHARAGI1995} S. K. Charagi, Phys. Rev. C{\bf 51} (1995) 3521.

\bibitem{Olmer1978} C. Olmer et.al, Phys. Rev.{\bf 18} (1978) 205.  
  
\bibitem{TMMEHIP} K. W. McVoy and W. A. Friedman, {\it Theoretical Methods in Medium-Energy and Heavy-Ion Physics}, 
  Springer US (1978).

\bibitem{PRC491652} Dao T. Khoa, W. von Oertzen, and H. G. Bohlen, Phys. Rev. C {\bf 49} (1994) 1652. 


\bibitem{PHYREVC26} M. Buenerd et. al, Phys. Rev. C{\bf 26} (1982) 1299.

\bibitem{NUCPHYA424} M. C. Mermaz, Nucl. Phys. A{\bf 424} (1984) 313.

\bibitem{NUCPHYA456} H. Ikezoe et. al, Nucl. Phys. A{\bf 456} (1986) 298.

\bibitem{NUOCIMA111} F. Nuoffer et. al, Nuovo Cimento A{\bf 111} (1998) 971.

\bibitem{PHYLETB365} G. Bartnitzky et. al, Phys. Lett. B{\bf 365} (1996) 23.

\bibitem{ALVAREZ} 
  M.~A.~G.~Alvarez, L.~C.~Chamon, M.~S.~Hussein, D.~Pereira, L.~R.~Gasques, E.~S.~Rossi, Jr. and C.~P.~Silva,
  Nucl.\ Phys.\ A {\bf 723} (2003) 93.
  [nucl-th/0210062].


  







\end{thebibliography}
\end{document}